\documentclass[aps,prl,twocolumn,showpacs,preprintnumbers,amsmath,amssymb,superscriptaddress]{revtex4}%

\usepackage{graphicx}
\usepackage{dcolumn}
\usepackage{bm}
\usepackage{color}
\usepackage{amssymb}

\begin{document}

\preprint{PREPRINT (\today)}

\newpage

\title{Pressure induced ferromagnetism in antiferromagnetic Fe$_{1.03}$Te}

\author{M.~Bendele}
\altaffiliation{Present address: Dipartimento di Fisica, Universit\`a di Roma ``La Sapienza'' - P. le Aldo Moro 2, I-00185 Roma, Italy}

\email{markus.bendele@physik.uzh.ch}
\affiliation{Physik-Institut der Universit\"{a}t Z\"{u}rich, Winterthurerstrasse 190, CH-8057 Z\"{u}rich, Switzerland}
\affiliation{Laboratory for Muon Spin Spectroscopy, Paul Scherrer Institute, CH-5232 Villigen PSI, Switzerland}

\author{A.~Maisuradze}
\affiliation{Physik-Institut der Universit\"{a}t Z\"{u}rich, Winterthurerstrasse 190, CH-8057 Z\"{u}rich, Switzerland}
\affiliation{Laboratory for Muon Spin Spectroscopy, Paul Scherrer Institute, CH-5232 Villigen PSI, Switzerland}

\author{B.~Roessli}
\affiliation{Laboratory for Neutron Scattering, Paul Scherrer Institute, CH-5232 Villigen PSI, Switzerland}

\author{S.~N.~Gvasaliya}
\affiliation{Neutron Scattering and Magnetism, Laboratorium f\"{u}r Festk\"{o}rperphysik, ETH Z\"{u}rich, CH-8093 Z\"{u}rich, Switzerland}

\author{E.~Pomjakushina}
\affiliation{Laboratory for Developments and Methods, Paul Scherrer Institute, CH-5232 Villigen PSI, Switzerland}

\author{S.~Weyeneth}
\affiliation{Physik-Institut der Universit\"{a}t Z\"{u}rich, Winterthurerstrasse 190, CH-8057 Z\"{u}rich, Switzerland}

\author{K.~Conder}
\affiliation{Laboratory for Developments and Methods, Paul Scherrer Institute, CH-5232 Villigen PSI, Switzerland}

\author{H.~Keller}
\affiliation{Physik-Institut der Universit\"{a}t Z\"{u}rich, Winterthurerstrasse 190, CH-8057 Z\"{u}rich, Switzerland}

\author{R.~Khasanov}
\affiliation{Laboratory for Muon Spin Spectroscopy, Paul Scherrer Institute, CH-5232 Villigen PSI, Switzerland}

\begin{abstract}
The magnetic properties of Fe$_{1.03}$Te under hydrostatic pressure up to $p\simeq 5.7$\,GPa were investigated by means of muon spin rotation, dc magnetization, and neutron depolarization measurements. With increasing pressure the antiferromagnetic ordering temperature $T_{\rm N}$ decreases continuously from 70\,K at ambient pressure towards higher pressures. Surprisingly, the commensurate antiferromagnetic order of FeTe enters a region of incommensurate and dynamical magnetic order before at $p\simeq 1.7$\,GPa the system turns ferromagnetic. The ferromagnetic ordering temperature $T_{\rm C}$ increases with increasing pressure.
\end{abstract}
\maketitle

By applying hydrostatic pressure the magnetic and superconducting properties of the iron-based superconductors can be directly controlled, since the carrier concentration and the exchange interaction through the compressed lattice are changed \cite{Torikachvili08,Takahashi08,Sun12,Khasanov11}. Within this new class of superconductors, the iron-chalcogenides have attracted considerable interest, because this system shows a large effect on the superconducting and magnetic properties upon both chemical and hydrostatic pressure. Substitution of Se by Te in superconducting FeSe leads to almost a doubling of the value of the superconducting transition temperature \emph{T}$_{\rm c}$ until the system enters a state of coexistence of superconductivity and magnetism, ending at antiferromagnetic FeTe \cite{Khasanov10}. On the other hand, application of hydrostatic pressure leads to a nonlinear increase of \emph{T}$_{\rm c}$ with a maximum value of $\simeq36$\,K \cite{Medvedev10}. Furthermore, magnetic order appears and superconductivity and magnetism coexist on a microscopic level in the whole sample \cite{Bendele10,Bendele12}. In antiferromagnetic FeTe a \emph{T}$_{\rm c}$ even higher than that of FeSe was predicted \cite{Subedi08,Zhang09,Mizugushi08}, but until now not observed \cite{Okada09}. Instead, the parent compound FeTe exhibits peculiar magnetic properties. At the N\'eel temperature $T_{\rm N}\simeq 70$\,K a drastic drop in the magnetic susceptibility is seen \cite{Li09,Bao09}. This is related to a first-order phase transition to a monoclinic crystal structure at $T_{\rm S}$ that coincides with the appearance of commensurate antiferromagnetism at $T_{\rm N}$ \cite{Furchart75}. However, at ambient pressure the magnetic properties of FeTe depend strongly on the amount of excess Fe \cite{Liu11,Rodriguez11}. For nearly stoichiometric Fe$_{1+x}$Te a distorted monoclinic structure and a commensurate antiferromagnetic order in the $ab$-plane was observed at low temperatures. In contrast, at a high amount of excess Fe the value of $T_{\rm N}$ decreases, and the magnetic order changes to an incommensurate antiferromagnetic one \cite{Liu11,Rodriguez11}. 

Pressure studies of the electrical resistivity up to 19\,GPa revealed that the anomaly in the resistivity at $T_{\rm S}$ shifts toward lower temperatures with increasing pressure \cite{Okada09}. Additionally, at high pressures a new anomaly in resistivity emerges, strongly indicating that the electronic properties of Fe$_{1+x}$Te are closely correlated with its crystal structure.

\begin{figure*}[htb]
\centering
\vspace{-0cm}
\includegraphics[width=\linewidth]{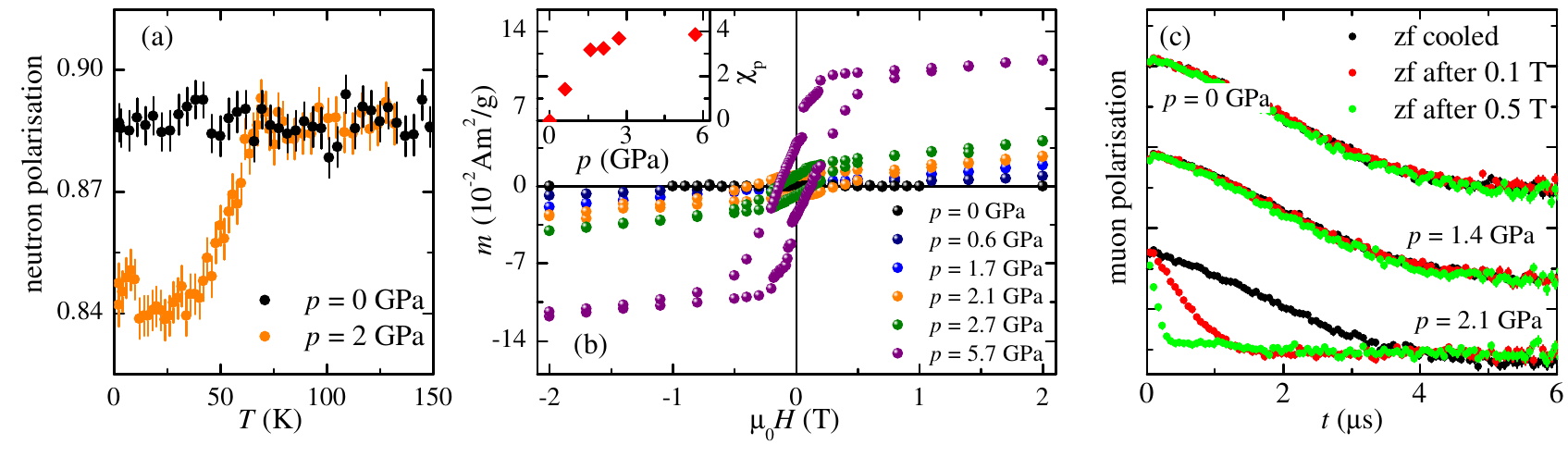}
\vspace{-1cm}
\caption{(color online) (a) Temperature dependence of the neutron polarization of Fe$_{1.03}$Te at ambient pressure and at $p=2$\,GPa. A clear loss of polarization is observed at $p=2$\,GPa below $T\simeq 60$\,K, indicating that the sample is ferromagnetic. (b) The measured hysteresis loops of Fe$_{1.03}$Te after subtraction of the ambient pressure data. Above $p\simeq1.7$\,GPa the for a ferromagnet typical opening of the hysteresis loop is evident indicating appearance of ferromagnetism in the sample under pressure. The inset in (b) shows the evolution of the paramagnetic susceptibility with increasing pressure. (c) zf $\mu$SR time spectra of Fe$_{1.03}$Te in a pressure cell at different pressures in zf cooled experiments and after application of $\mu_0H=0.1$ and $0.5$\,T. At high pressures ($p\geq1.7$\,GPa) a difference in the time spectra is seen, indicating the appearance of a net-magnetic moment which signals the presence of a ferromagnetic phase in the sample. }
\end{figure*}

In this work, a systematic investigation of the magnetic properties of Fe$_{1.03}$Te under hydrostatic pressure using neutron depolarization, magnetization, and muon spin rotation ($\mu$SR) experiments is reported. All the three experimental techniques reveal consistently a transition from a low pressure antiferromagnetic phase to a high pressure ferromagnetic phase in Fe$_{1.03}$Te (see Fig. 1). The neutron depolarization experiments were performed at the Paul Scherrer Institute (PSI) using the TASP 3-axis spectromenter equipped with a MuPAD polarimeter \cite{tasp,mupad}.
Figure 1a presents data of the polarized neutron experiments taken on Fe$_{1.03}$Te in a piston cylinder cell with a neutron wavelength of $3.2$\,\AA. In these measurements the polarization of a monochromatic neutron beam is measured after transmission through the sample. Because of the Larmor precession, the polarization of the beam will rotate around the magnetic fields present in the sample. If the sample contains randomly aligned ferromagnetic domains, the polarization vector will rotate around different directions resulting in a loss of initial polarization (see Fig. 1a). The temperature dependence of the neutron polarization at ambient pressure shows no loss of polarization, whereas at $p\simeq 2$\,GPa the neutrons loose $\sim5$\% of their polarization below the ferromagnetic ordering temperature $T_{\rm C}(2\,{\rm GPa})\simeq 60$\,K.  

The ferromagnetic behavior of Fe$_{1.03}$Te under pressure is further evidenced by field dependent magnetization measurements in a commercial {\it Quantum Design} 7 T Magnetic Property Measurement System (MPMS) XL SQUID magnetometer. The measurements up to $p\simeq 6$\,GPa were performed in a diamond anvil cell, especially designed for magnetization measurements \cite{pressurecell}. The field dependencies of the magnetization after subtraction of the zero pressure measurements are shown in Fig. 1b. The mass of the Fe$_{1.03}$Te sample in the pressure cell was determined by normalizing the drop in the susceptibility at 70\,K measured in the pressure cell at ambient pressure to a large sample measured without pressure cell resulting in $m\simeq0.09$\,mg. The typical hysteretic behavior of a ferromagnet is obvious above $p\simeq1.7$\,GPa. At higher pressure this behavior is more pronounced. However, the coercive field seems to be maximal with a value of 0.35\,T at $p\approx 2$\,GPa. At even higher pressures the value saturates at $\sim0.15$\,T. 

Furthermore, the change of the paramagnetic susceptibility compared to zero pressure $\chi_{\rm p}(p)$ can be extracted from these measurements by fitting a linear slope to the saturated magnetization following the simple relation $m=\chi_pH$. It increases with increasing pressure (see inset Fig. 1b) in agreement with an earlier work \cite{Fedorchenko11}. This is attributed to the itinerant magnetism of Fe$_{1+x}$Te. As soon as Fe$_{1.03}$ becomes ferromagnetic $\chi_{\rm p}$ tends to saturate with increasing pressure which is expected for ferromagnetic ordering. 


The $\mu$SR experiments were performed at PSI on the $\mu$E1 beam line at the GPD instrument. In a $\mu$SR experiment, spin polarized muons are implanted into the sample and by monitoring the time evolution of the muon spin polarization, information on the local magnetic field at the muon stopping site $B_\mu$ and the magnetic volume fraction $F$ are obtained. Here, the internal magnetic field distribution of Fe$_{1.03}$Te was investigated for different pressures by means of zero field (zf) $\mu$SR at 10\,K. The signal consists of a superposition of the one arising from the sample and the one from the pressure cell that is pressure independent. In fact, with the help of the pressure cell it can be directly determined, whether the sample is antiferromagnetic or ferromagnetic: as soon as Fe$_{1+y}$Te orders ferromagnetically a net magnetization remains after a magnetic field was applied (remanent magnetic field). As a result, the zf $\mu$SR signals of the pressure cell before and after the application of a magnetic field are different. In Fig.~1c it is seen that at low pressures up to 1.6\,GPa the $\mu$SR time spectra overlap before and after application of a magnetic field. Thus, the sample is in the antiferromagnetic state. However, above $p\simeq1.8$\,GPa the signals clearly differ from each other after application of a magnetic field, indicating that antiferromagnetic Fe$_{1.03}$Te becomes ferromagnetic under pressure.

\begin{figure}[tb]
\centering
\vspace{-0cm}
\includegraphics[width=0.9\linewidth]{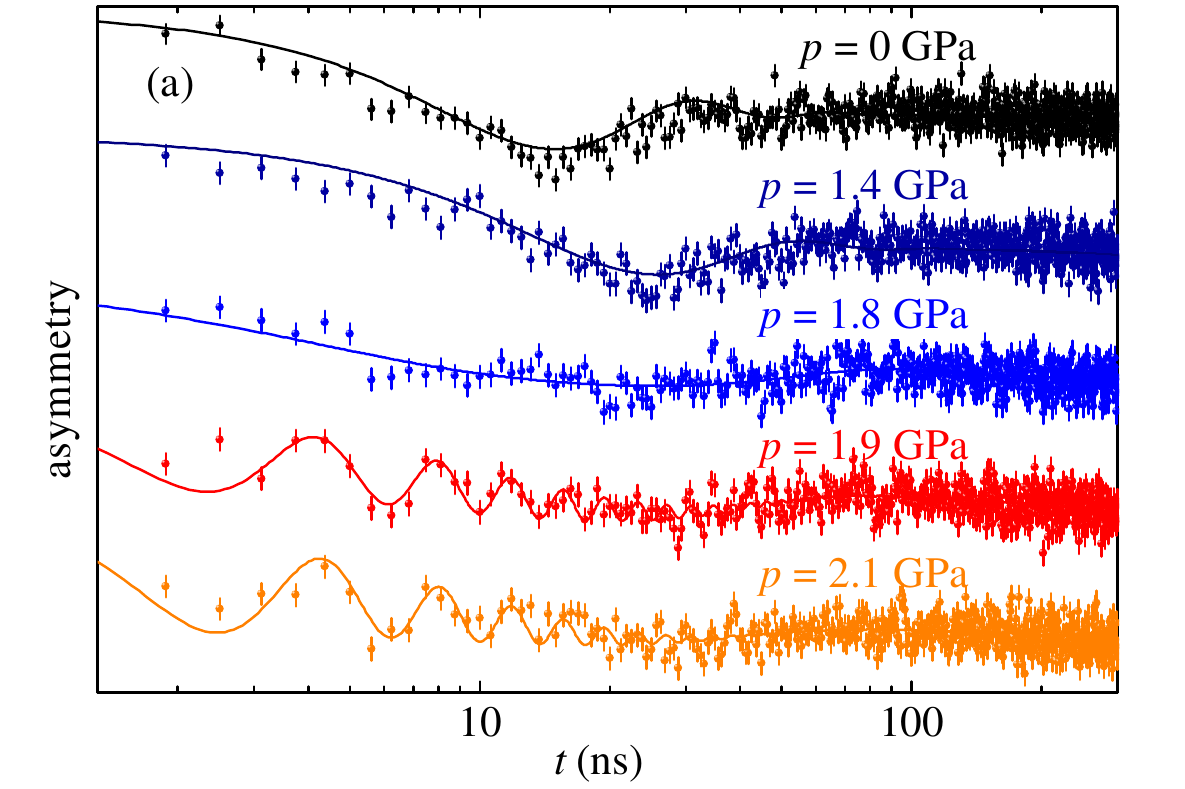}\\\vspace{-0.5cm}
\includegraphics[width=0.8\linewidth]{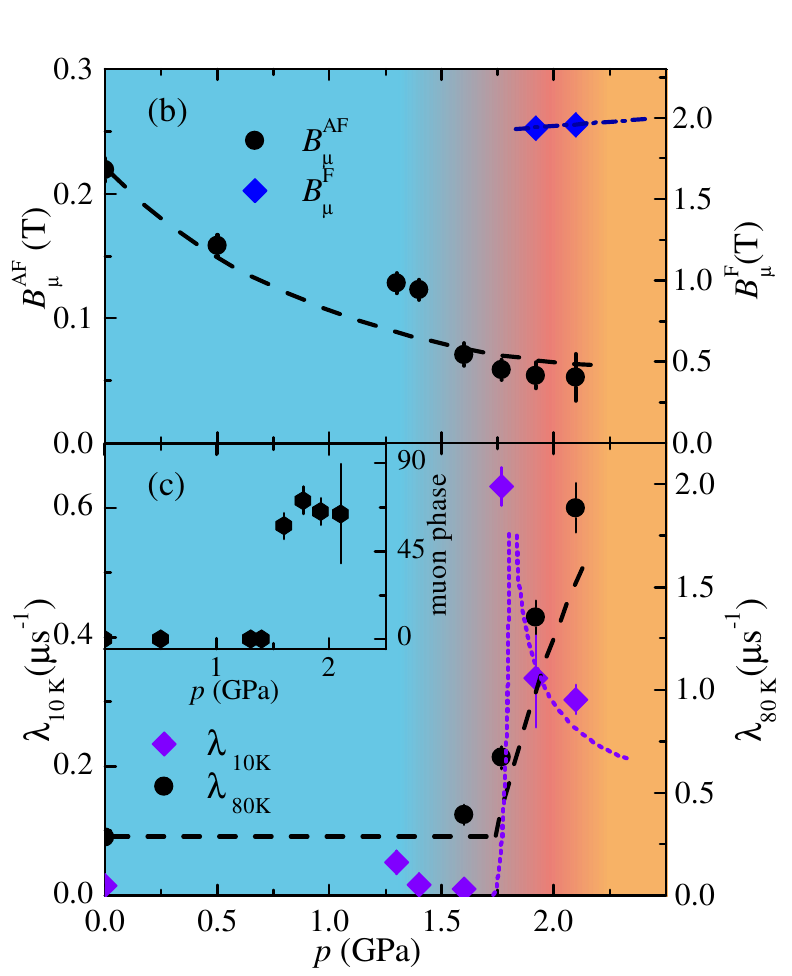}
\vspace{-0.5cm}
\caption{(color online) (a) zf $\mu$SR time spectra of Fe$_{1.03}$Te for various pressures at $T=10$\,K. Above $p\simeq1.6$\,GPa a relaxation at high times is observed pointing to the appearance of dynamical order. At high pressures $p\geq1.9$\,GPa an additional high local magnetic field $B^{\rm F}_{\mu}\simeq1.9$\,T appears in the spectra. This is assigned to the appearance of ferromagnetic order.
(b) Pressure dependence of the internal magnetic fields $B_{\mu}^{\rm AF}$ (left axis) and $B_{\mu}^{\rm F}$ (right axis). (c) Pressure dependence of the longitudinal relaxation rates $\lambda_{10\,{\rm K}}$ at $10$\,K (left axis) and $\lambda_{80\,{\rm K}}$ at $80$\,K (right axis) observed at high times [see (a)]. In the vicinity of the ferromagnetic state $\lambda_{10\,{\rm K}}$ has a maximum and tends to decrease again as soon as ferromagnetism is established.  Above $p\simeq1.7$\,GPa the relaxation rate $\lambda_{80\,{\rm K}}$ increases strongly indicating fluctuations. The inset shows the pressure dependence of the initial muon phase. At $p\simeq 1.6$\,GPa a step of the phase is observed.
}
\end{figure}

A more detailed view on the $\mu$SR signal of the Fe$_{1.03}$Te sample presented in Fig. 2a shows a spontaneous precession of the muon spins, indicating a long-range ordered magnetic state in the sample. For all pressures up to $p\lesssim 1.6$\,GPa a model function describing commensurate magnetic order was used \cite{Khasanov10}, in agreement with a recent neutron study \cite{Jorgensen12}. The muon precession frequency extracted from the $\mu$SR spectra is directly proportional to the local magnetic field $B_{\mu}^{\rm AF/F}$ at the muon stopping site, where AF and F refer to the antiferromagnetic and ferromagnetic state, resepectively. The pressure dependence of $B_{\mu}^{\rm AF}$ is shown in Fig. 2b. It decreases with increasing pressure, whereas it tends to saturate at high pressures.
For higher pressures close to the ferromagnetic state ($p\sim 1.7$\,GPa) the magnetic order changes to a more complicated one that is incommensurate and dynamical. This is concluded, since it is necessary to introduce an additional phase parameter to the model (see inset of Fig. 2c). 
The dynamical character of the magnetism in the crossover region from antiferromagnetism to ferromagnetism is demonstrated by the steep increase of the longitudinal relaxation rate $\lambda_{10\,{\rm K}}$ at 10\,K with increasing pressure shown in Fig. 2c. 
Typical for static magnetism it is almost zero at low pressures, whereas in the crossover region $\lambda_{10\,{\rm K}}$ increases substantially. In the ferromagnetic state $\lambda_{10\,{\rm K}}$ tends to decrease again, indicating that the magnetic order becomes static at high pressures. 
The same effect of an increased relaxation above $p\gtrsim1.6$\,GP is observed at high temperatures. However, in contrary to low temperatures the relaxation further increases with increasing pressure in the ferromagnetic phase as for example demonstrated by $\lambda_{80\,{\rm K}}$ at 80\,K shown in Fig. 2c. This faster relaxation at high temperatures suggests that the increasing fluctuations are related to the appearance of pressure-induced ferromagnetism. 
This, however, needs further detailed investigations.
In the ferromagnetic state ($p\gtrsim 1.9$\,GPa) an additional high field of $B_{\mu}^{\rm F}\gtrsim 1.7$\,T is observed in the $\mu$SR time spectra which is allocated to the ferromagnetic moment and is substantially larger than $B_{\mu}^{\rm AF}$ (see Fig. 2a). Therefore, a second model function for a magnetic order was introduced in order to take $B_{\mu}^{\rm F}$ into account leading to a superposition of the two oscillations. However, $B_{\mu}^{\rm AF}$ remains present till at least $p\simeq 2.1$\,GPa, but with a decreasing volume fraction.

\begin{figure}[tb]
\centering
\vspace{-0cm}
\includegraphics[width=\linewidth]{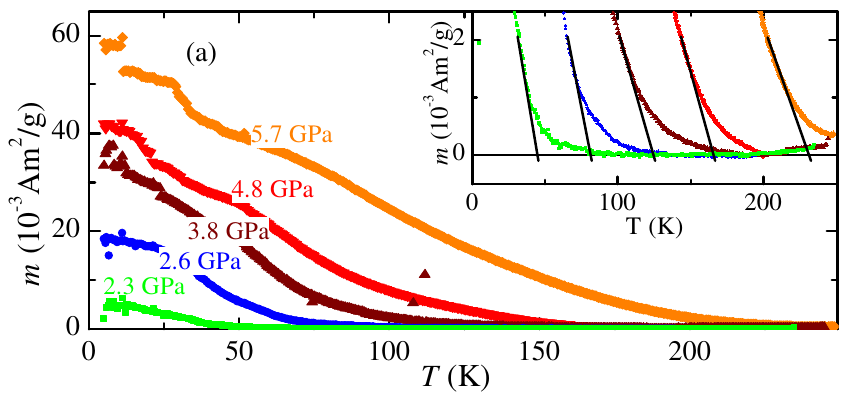}\\ 
\includegraphics[width=\linewidth]{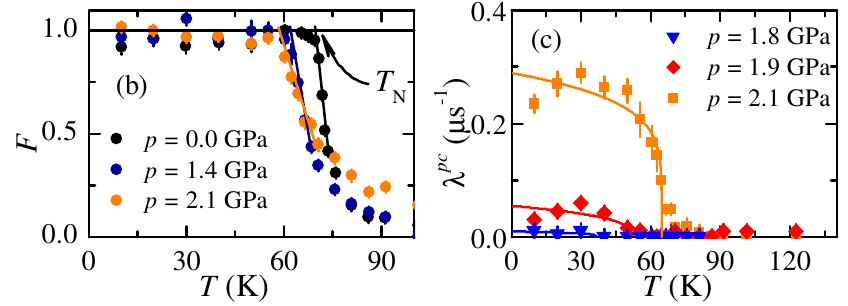}
\vspace{-0.5cm}
\caption{(color online) (a) Temperature dependence of the magnetic moment $m$ of Fe$_{1.03}$Te at $\mu_0H=0.1$\,T after subtraction of the zero pressure data for various pressures. The inset shows the determination of $T_{\rm C}$ by linear extrapolation of $m$ to zero.
(b) Temperature dependence of the magnetic fraction $F$ for selected pressures. 
(c) Temperature dependence of the additional relaxation $\lambda^{pc}$ of the pressure cell due to the ferromagnetic state of the sample.
 }
\end{figure}

The evaluation of the ferromagnetic state was further investigated by means of temperature dependent magnetization measurements in a magnetic field of $\mu_0H=0.1$\,T shown in Fig. 3a. Similar to the field dependencies, the ambient pressure measurements were subtracted in order to extract the ferromagnetic signal of the sample. In addition, the paramagnetic contribution above $T_{\rm C}$ of the 2.3\,GPa data was fitted to a $m\propto aT+C/T$ behavior and subtracted for all pressures. The Curie temperature $T_{\rm C}$ was determined from the intersection of the linear extrapolation of the magnetic moment close to $T_{\rm C}$ with the zero line (see inset of Fig. 3a). 

Both the ferromagnetic ($T_{\rm C}$) and the antiferromagnetic ($T_{\rm N}$) transition temperatures were obtained by means of weak transverse field $\mu$SR measurements (for details see Ref. \cite{Khasanov10}). 
From these measurements, the temperature dependence of the magnetic fraction $F$ for various pressures is obtained (see Fig. 3c). Again, $T_{\rm N}$ is obtained by the intersection of the linear extrapolation of the data points close to the transition with the line representing 100\% magnetic volume fraction. 
Furthermore, the ferromagnetic ordering temperature $T_{\rm C}$ can also be determined from these measurements with the help of the pressure cell signal. Once Fe$_{1.03}$Te is in the ferromagnetic state, it is polarized in a magnetic field and has a net magnetization with a stray field. This leads to an increase of the field inhomogeneities in the pressure cell, and consequently the signal of the pressure cell becomes pressure dependent (see Fig. 3c). 

\begin{figure}[t]
\centering
\vspace{-0.5cm}
\includegraphics[width=\linewidth]{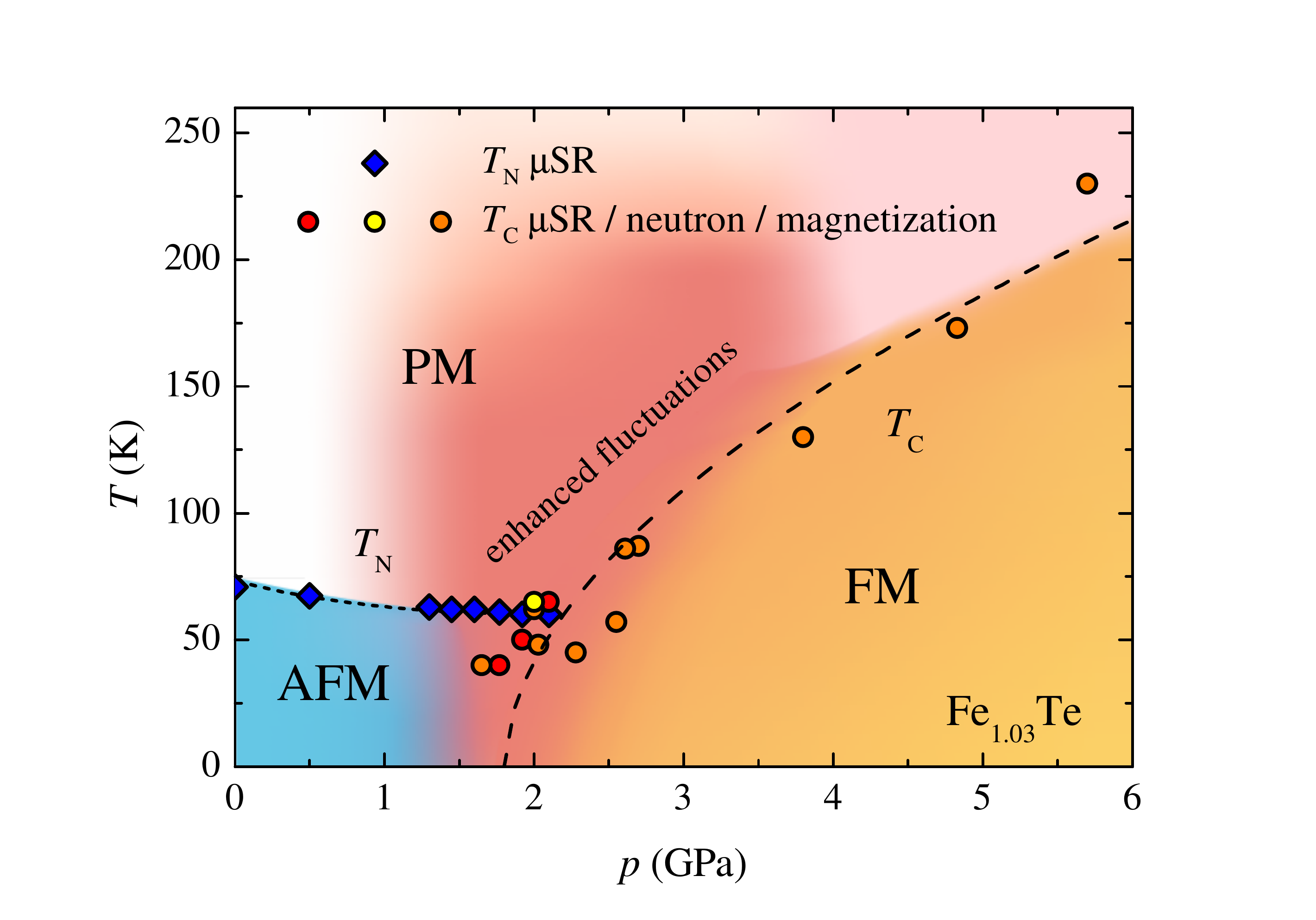}
\vspace{-1cm}
\caption{(color online) Pressure dependence of the magnetic ordering temperatures $T_{\rm N}$ and $T_{\rm C}$ of Fe$_{1.03}$Te. At $p\simeq1.6$\,GPa the commensurate antiferromagnetic order (AFM) changes to an incommensurate, dynamical magnetic order. Above $p\simeq1.7$\,GPa Fe$_{1.03}$Te becomes ferromagnetic (FM). In the paramagnetic state (PM) fluctuations appear in vicinity of the ferromagnetic state. The dashed lines indicating $T_{\rm N}(p)$ and $T_{\rm C}(p)$ are guides to the eyes. 
}
\end{figure}

The values of $T_{\rm C}$ and $T_{\rm N}$ extracted from the neutron, magnetization, and $\mu$SR measurements are summarized in a phase diagram exhibiting different magnetic states (see Fig.~4). At ambient pressure Fe$_{1+x}$Te exhibits commensurate antiferromagnetic ordering along the $(\pi , 0)$ direction \cite{Bao09}. The magnetic transition at $T_{\rm N}$ is accompanied by a first-order structural phase transition.  Application of pressure first leads to a decrease of $T_{\rm N}$ and a reduction of $B_{\mu}^{\rm AF}$. At high pressure the commensurate antiferromagnetic order of Fe$_{1.03}$Te turns into incommensurate and dynamical magnetic order, in agreement with a recent neutron study \cite{Jorgensen12}. Surprisingly, after further increase of hydrostatic pressure above a relatively low value of $\simeq1.7$\,GPa the system becomes ferromagnetic. Thus, the observed dynamical and incommensurate order before the occurrence of ferromagnetism could be non-orientated spin clusters since in general in crossover regions from an antiferromagnet to a ferromagnet such a behavior can be observed.
A change in the crystal lattice cannot be responsible for this behavior, because at room temperature the structure is tetragonal up to 4\,GPa. At higher pressures, however, a collapsed tetragonal structure was observed \cite{Zhang09_chrystal}. Note that different pressure induced magnetic phases were already suggested in an earlier work, however, they could not be identified \cite{Okada09}. 

An antiferromagnet turning into a ferromagnet by applying hydrostatic pressure is to the best of our knowledge a unique feature, since in general the opposite is observed \cite{Ding09,Umeo10}. In comparison, an antiferromagnet can be easily tuned into a ferromagnet by application of a high  magnetic field \cite{Weyeneth11,Guguchia11}. However, in this case not an itinerant antiferromagnetic spin density wave, but localized moments order ferromagnetically in these compounds. On the other hand, Fe$_{1+x}$Te is an itinerant antiferromagnet without rare earth atoms. Yet, a density functional study has shown that the interstitial iron acts as a strong local moment \cite{Zhang09}, that could order ferromagnetically. However, introducing more interstitial iron does not lead to ferromagnetism in the system \cite{Rodriguez11}.
Another theoretical model describing the magnetic order in Fe$_{1-x}$Te suggests the commensurate ordered material to be in vicinity of a incommensurate antiferromagnetic and a ferromagnetic phase \cite{Fang09}. 

In conclusion, antiferromagnetic Fe$_{1.03}$Te shows pressure induced ferromagnetism above $p=1.7$\,GPa after crossing a region of dynamical and incommensurate antiferromagnetism. 
This peculiar observation may help to unravel the complex magnetic and superconducting properties in Fe-based systems.

Helpful discussions with R. Puzniak are greatfully acknowledged. This work was supported by the Swiss National Science Foundation and the NCCR program MaNEP. The experiments were partially performed at the Swiss Muon Source S$\mu$S and at the Swiss spallation neutron source SINQ of the Paul Scherrer Institute, PSI Villigen, Switzerland. The $\mu$SR data were analyzed using the free software package MUSRFIT \cite{MusrFit}.

\end{document}